\documentclass{ws-mpla}
\usepackage[super]{cite}
\usepackage{graphicx}
\usepackage[utf8]{inputenc}
\usepackage[T1]{fontenc}
\usepackage{lmodern}
\usepackage[a4paper]{geometry}
\usepackage{amsmath, amssymb}
\usepackage{xcolor}
\usepackage[french]{babel}

\usepackage{comment}
\includecomment{prof}

\begin{document}

\markboth{Authors' Names}
{Instructions for Typing Manuscripts (Paper's Title)}

\catchline{}{}{}{}{}

\title{Thermodynamic Phase Transition and global stability of the Regular Hayward Black hole Surrounded by Quintessence \\
\footnote{For the title, try not to use more than
three lines. Typeset the title in 10 pt Times Roman, uppercase and
boldface.}
}

\author{\footnotesize Kamiko Kouemeni Jean Rodrigue}

\address{Department of Physics, Faculty of Science, University
of Maroua, P.O. Box. 814 Maroua, Cameroon}

\author{Mahamat Saleh}

\address{Department of Physics, Higher Teacher's Training College, University
of Maroua, P.O. Box. 46, Cameroon.}

\author{Bouetou Bouetou Thomas}

\address{Ecole Nationale Sup$\acute{e}$rieure Polytechnique,
University of Yaounde I, P.O. Box. 8390, Cameroon.\\ The African Center of Excellence in
 Information and Communication Technologies (CETIC),Yaounde, Cameroon. \\ and}

\author{Kofane Timoleon Crepin}

\address{Department of Physics, Faculty of Science, University
of Yaounde I, P.O. Box. 812, Cameroon. \\ National Advanced School of Engineering, University of
Yaounde I, P.O. Box 8390, Yaounde,Cameroon.}

\maketitle

\pub{Received (Day Month Year)}{Revised (Day Month Year)}

\begin{abstract}
In this work, we investigate the thermodynamic and the stability of the regular Hayward black hole surrounded by quintessence. Using the metric of the black hole surrounded by quintessence and the new approach of the holographic principle, we derive the expression of the Unruh Verlinde temperature. Hawking temperature and specific heat are derived using the first law of black holes thermodynamics. Gibbs free energy is also evaluated. The behaviors of these quantities show that, the quantum effects represented by the parameter $\beta$ induces a decreasing of the Hawking temperature of the black hole, and that decrease is accentuated when increasing the magnitude of $\beta$ and the normalization factor $a$ related to the density of quintessence. For the lower entropies, the black hole passes from the unstable phase to the stable one by a first order thermodynamics phase transition. When increasing the entropy, a second phase transition occurs. This new phase transition is a second-order thermodynamics phase transition and brings the black hole to unstable state. It results that, when increasing of magnitude of $\beta$, the phase transition points are shifted to the higher entropies. Moreover, the phenomena of phase transitions are preserved by adding the quintessence. Furthermore, when increasing the normalization factor of quintessence, the first order transition point is shifted to higher entropies, while the second-order thermodynamics phase transition point is shifted to lower entropies.

\end{abstract}

\ccode{PACS Number: 04.70.Bw, 04. 20 Jb, 95.36.+x , 04.70.Dy.}

\section{Introduction}	

\label{intro}
The solution to Einstein's field equations discovered by Schwarzschild in $1916$ \cite{R01,R1}, has been the starting for black hole physics. Following this solution, many scientists have investigated more and more in this field and found other solutions describing charged and rotating black holes \cite{R1,R02,R2}, as well as the perturbed black holes \cite{R2,R3,R4,R5,R6,R7,R8,R9}. However, a major problem, namely the singularity, which is the famous point where all the physical quantities which describe the space-time curvature become infinite, characterizes these solutions. This problem of black hole singularity is an acknowledged difficulty in general relativity \cite{R09,R10,R11,R12,R1201,R14}. To solve this problem, scientists decide to construct the solutions of Einstein's equation where the metric and the curvature invariants are all regular everywhere \cite{R12,R1201}. The first of these solutions called regular black hole and which does not contain a singularity at the center has been constructed by Bardeen in $1968$ \cite{R09,R11,R12,R13,R16}. In $2000$ Ayón-Beato and A. García proved that the Bardeen black hole was an exact solution in a model of spacetime with nonlinear electrodynamics \cite{R1601}. The action of nonlinear electrodynamics in curve spacetime is given by
\begin{equation}\label{eq1}
    S=\frac{1}{16\pi}\int d^{4}x\sqrt{-g}[R-\emph{L}(F)],
\end{equation}
where $R$ is the scalar curvature and the $\emph{L}$ depends on the $\emph{F}=F_{\mu\nu}F^{\mu\nu}$. After Bardeen solutions, regular black holes are more investigated and different other solutions like Hayward black hole \cite{R17}, and modified Hayward black hole \cite{R13,R17,R18} were constructed.

The works of Hubble on the universe observations shown that our universe is growing with time. But the question of the slow down of the rate of that growing due to the influence of gravity remains until that some astronomers like Walter Baade and Fritz Zwicky explored the super-novae phenomenon. The recent measurement of these studies prove that our universe is expanding too fast \cite{R108,R109,R1010}. Recently in $2011$, the crucial importance of this phenomenon has earned to physicists Saul Perlmutter, Brian P. Schmidt and Adam G. Riess the Nobel Prize for physics \cite{R109,R1010}. So,at the question what will be the final destiny of the Universe: Some say the world will end in fire, some say in ice? For us, Probably it will end in ice, if we are to believe this year’s Nobel Laureates in Physics. They have studied several dozen exploding stars, called super-novae, and discovered that the Universe is expanding at an ever-accelerating rate. The discovery came as a complete surprise even to the Laureates themselves. Through the high-precision observational data, it is well known that the accelerating expansion of the universe is due to the mysterious form of energy called dark energy \cite{Re000,Re00,Re0,Re1,Re2,Re201,Re3,Re4,Re5,Re6,Re701,Re702,Re703}. Like cosmological constant \cite{Ri7,Ri8}, more other candidates of dark energy described by an equation of state $p=\rho_{q} \omega_{q}$ are investigated today \cite{Ri9,Ri10,Ri11,Ri1101,Ri12,Ri13,Ri14,Ri15,Ri16,Ri17,Ri18,Ri19}. Quintessence is a dark energy candidate for which the state parameter $\omega_{q}$ is in the range of $-1\le \omega_{q}\le -1/3$ \cite{R4,Ri20}. The contradictory effects of a black hole and dark energy on the future of the universe have pushed scientists to study the consequences of their co-existence in our universe. Studying the influence of dark energy on the behaviors of a black hole is actually investigated by more scientists \cite{R4,Ri20,Re7,Re8,Re9,Re10,Re11,Re12}.

In $1974$, the physicists Hawking and Bekenstein showed from semiclassical point, that black holes emit radiation. In this consideration, black hole is presented as thermodynamic object \cite{Rt1,Rt2}. Through this issues, different thermodynamics quantities were founded to the describe black hole thermodynamically. Hawking personally showed that, the temperature of a black hole at the event horizon could be expressed as $T_{H}=\frac{\partial E}{\partial S}$, where $E$ is the internal black hole energy and $S=\pi r_{H}^2$ is Bekenstein-Hawking entropy of black hole  at the event horizon \cite{R4}. Unruh-Verlinde temperature \cite{R1,R3} and Tolman temperature \cite{Rt3} generalize Hawking temperature at any point of inside or outside black hole. Nowadays in black hole physics, the thermodynamics phase transition is one of most important phenomena which is used to investigate their stabilities. Black hole thermodynamics phase transition can be studied using Hawking temperature \cite{Rik20,Rik21}, specific heat \cite{R4,Ri20,Re9,Rik22,Rik23}, free energy \cite{Rt4,Rt5}, and Gibbs free energy \cite{Rik22,Rt05}. These different quantities can be investigated as the function of black hole mass \cite{Rik20,Rik21}, horizon size \cite{Re9} or entropy \cite{R4,Ri20,Rik22,Rik23}.

Recently, Mehdipour et \emph{al.} \cite{Rso1} have investigated thermodynamics and phase transition of Hayward solutions. More recently, Saleh et \emph{al.} \cite{Rim02} have studied the effects of quintessence on the thermodynamics and phase transition from the regular
Bardeen black hole. In this paper, we aim at investigating the effects of quintessence on the thermodynamics behavior of Hayward black hole.

The paper is organized as follows. In Sec. \ref{sec:a}, we express the metric of the Hayward black hole with quintessence. In Sec. \ref{sec:b}, we derive the expression of the Unruh-Verlinde temperature and show the effects of quintessence on the temperature of regular Hayward black hole. In Sec. \ref{sec:c}, we derive the expression of the Hawking temperature and study its behavior with respect to the quintessence normalization factor and the parameter of regular Hayward black hole. In Sec. \ref{sec:d}, we establish the expression of the specific heat of the black hole, afterward we show the effects occurring due to the presence of quintessence. In Sec. \ref{sec:cg} we derive the expression of Gibbs free energy using the classical definition of the enthalpy, and shows the effects of $\beta$ and the normalization factor of quintessence dark energy on its behaviors. Sec. \ref{sec:e} is reserved to the conclusion.


\section{Expression of the metric of Hayward black hole surrounded  by quintessence}
\label{sec:a}

The metric of Hayward regular black hole is given by \cite{R09,R10}
\begin{equation}\label{em1}
    dS^{^2}=-f(r)dt^{2}+f(r)^{-1}dr^{2}+r^{2}d\theta^{2}+r^{2}\sin^{2}\theta d\varphi^{2},
\end{equation}
where
\begin{equation}\label{em2}
    f(r)=1-\frac{2M(r)}{r} ,
\end{equation}

with $M(r)=\frac{mr^{3}}{r^{3}+2\beta^{2}}$, $\beta$ represents the correction term due to the quantum effects \cite{R09,Rk09,Rk091}, Indeed, Hayward stepped that, $\beta^{2}=ml^{2}$, where $l$ is expected to be the planck length \cite{R09} and $m$ is the black hole mass.

Kiselev proposed a new solution of the metric for static and spherically symmetric space-time in the presence of quintessence \cite{Re9}. Through that work, it results that the expression of the metric function of such black hole surrounded by quintessence is obtained by adding the term $-\frac{a}{r^{3\omega_{q}+1}}$ to that of the black hole free from quintessence \cite{Re9,Rso3}. $\omega_{q}$ is the state parameter of quintessence dark energy ranges between $-1$ and $-\frac{1}{3}$ and $a$ represents the normalization factor related to density of quintessence $\rho_{q}$ by the relation $\rho_{q}=-\frac{a}{2}\frac{3\omega_{q} }{r^{3(\omega_{q}+1)}}$. Actually, the values o the normalization factor is assumed to be positive and very low \cite{Re9,Rek09,R4}.

With these considerations, the Hayward black hole metric  surrounded by quintessence would be deduced from the metric of the regular Hayward black hole by adding the term $-\frac{a}{r^{3\omega_{q}+1}}$ on Eq.(\ref{em2}) as follows
\begin{equation}\label{em8}
    dS^{^2}=-f(r)dt^{2}+f(r)^{-1}dr^{2}+r^{2}d\theta^{2}+r^{2}\sin^{2}\theta d\varphi^{2},
\end{equation}
with
\begin{equation}\label{em9}
    f(r)=1-\frac{2M(r)}{r}-\frac{a}{r^{3\omega_{q}+1}}.
\end{equation}

Using this solution, we are going to investigate the thermodynamics behaviors of the Hayward black hole surrounded by quintessence dark energy.
\section{Unruh-Verlinde temperature of Hayward black hole with quintessence}
\label{sec:b}

Considering a static background with a global time like Killing vector $\xi^{\alpha}$, Verlinde derived the expressions of the potential $\phi$ and temperature $T$ in the form \cite{R1,R3,Rko1}
\begin{equation}\label{eq4}
    \phi=\frac{1}{2}log\Big(-g^{\alpha\beta}\xi_{\alpha}\xi_{\beta}\Big),
\end{equation}

\begin{equation}\label{eq5}
    T=\frac{\hbar}{2\pi} e^{\phi}n^{\alpha}\nabla_{\alpha}\phi,
\end{equation}
where a red-shift factor $e^{\phi}$ is inserted because the temperature $T$ is measured with respect to the reference point at infinity. The quantity $e^{\phi}$ is  supposed to be equal to unity at the infinity ($\phi= 0 \ at\ r = \infty$), if the space-time is asymptotically flat. $n$ is the number of bits on screen. The temperature defined in  Eq.(\ref{eq5}) is called the Unruh-Verlinde temperature.

Using the natural unit, these expressions above for a static and spherically symmetric black holes are reduced to \cite{R1}:
\begin{equation}\label{eq6}
    \phi=\frac{1}{2}\log(f(r)),
\end{equation}

\begin{equation}\label{eq7}
   T=\frac{1}{4\pi}\mid f'(r)\mid,
\end{equation}
where the symbol prime $(')$ represents the derivative with respect to the radial coordinate, $r$.

Substituting Eq.(\ref{em8}) and Eq.(\ref{em9}) in Eq.(\ref{eq7}), the expression of the Unruh-Verlinde temperature for the regular
Hayward black hole surrounded by quintessence is
\begin{equation}\label{tuh}
   T=\frac{mr(r^3-4\beta^2)}{2\pi(r^3+2\beta^2)^2}+\frac{(3\omega_{q}+1)a}{4\pi r^(3\omega_{q}+2)}.
\end{equation}
Considering only the quantum effects ($\beta\neq0$ and $a=0$), this expression of the Unruh-Verlinde temperature is reduced to
\begin{equation}\label{tuh0}
    T=\frac{mr(r^3-4\beta^2)}{2\pi(r^3+2\beta^2)^2}.
\end{equation}

Taking account the presence of quintessence $(a\neq0)$ but neglecting the correction due to the quantum effects $(\beta=0)$, the expression of the Unruh-Verlinde temperature becomes
\begin{equation}\label{tuh2}
  T=\frac{m}{2\pi r^2}+\frac{(3\omega_{q}+1)a}{4\pi r^{(3\omega_{q}+2)}}.
\end{equation}
This expression corresponds to the Unruh-Verlinde temperature of the Schwarzschild black hole surrounded by quintessence.
Setting $a=0$, this expression becomes
\begin{equation}\label{tuh1}
    T=\frac{m}{2\pi r^2},
\end{equation}
which is the Unruh-Verlinde temperature obtained by Liu et \emph{al.} \cite{R1} for Schwarzschild black hole without any perturbation field .

\section{Hawking temperature of regular Hayward black hole with quintessence}
\label{sec:c}
 At the event horizon, the metric function $f(r)$ vanishes and the radius of event horizon $(r_{h})$ is obtained by
\begin{equation}\label{th0}
     f(r_{h})=1-\frac{2M(r_{h})}{r_{h}}-\frac{a}{r_{h}^{3(\omega_{q}+1)}}=0.
\end{equation}
This equation leads to
\begin{equation}\label{th1}
    m=\frac{(r^{(3\omega_{q}+1)}_{h}-a)(r^{3}_{h}+2\beta^{2})}{2r^{(3\omega_{q}+3)}}.
\end{equation}
Using the area law, the black hole entropy $S$ is expressed as
\begin{equation}\label{th2}
    S=\pi r_{h}^{2}.
\end{equation}
From Eq.(\ref{th1}) and Eq.(\ref{th2}), we obtain the expression of black hole mass as a function of the entropy $S$, normalization factor of quintessence $a$ and quantum correction term $\beta$ \cite{R12,Rik23}
\begin{equation}\label{th3}
    m(S,a,\beta)=\frac{\pi}{2}\frac{\Big((\frac{S}{\pi})^{\frac{(3\omega_{q}+1)}{2}}-a\Big)\Big((\frac{S}{\pi})^{\frac{3}{2}}+2\beta^{2}\Big)}{2S(\frac{S}{\pi})^{\frac{(3\omega_{q}+1)}{2}}}.
\end{equation}

looking that expression of the black hole mass $m(S,a,\beta)$, the first law of lack hole thermodynamics can takes the form
\begin{equation}\label{eqcor1jmpa01}
dE=TdS+V_{a}da+V_{\beta}d\beta,
\end{equation}
\\where $E$ is the internal energy of black hole. However for the regular black holes, $dE$ is related to the black hole mass $dm$ by the corrected form of the first law of lack hole thermodynamics \cite{R12,Rim02}
\begin{equation}\label{th5}
   \delta E=C(m,r_{h})\delta m,
\end{equation}
with $C(m,r_{h})=\frac{M(r_{h})}{m}$. \\The others quantities derived by
\begin{equation}\label{th4}
T_{H}=\Big(\frac{\partial E}{\partial S}\Big)_{a,\beta}
\end{equation}
\begin{equation}\label{eqcor1jmpa02}
V_{a}=\Big(\frac{\partial E}{\partial a}\Big)_{S,\beta}
\end{equation}
\begin{equation}\label{eqcor1jmpa03}
V_{\beta}=\Big(\frac{\partial E}{\partial \beta}\Big)_{S,a}
\end{equation}
 represent respectively, the Hawking temperature, the thermodynamic potential induced by the quintessence dark energy and the thermodynamics potential induced by the quantum effects of the black hole.

 Substituting Eq.(\ref{th3}) and Eq.(\ref{th5}) in Eq.(\ref{th4}), the Hawking temperature for Hayward regular black hole surrounded by quintessence is obtained in the form
\begin{equation}\label{th6}
\begin{array}{ll}
   T_{H}=& \frac{1}{4S\big(S\sqrt{\frac{S}{\pi}}+2\pi\beta^2\big)}\Big(3a\big(\omega_{q} S\sqrt{\frac{S}{\pi}}+2\pi\omega_{q}\beta^2+2\pi\beta^2\big)\big(\frac{S}{\pi}\big)^{-\frac{3\omega_{q}}{2}}\\
   & +\frac{S^2-4\pi^2\beta^2}{\pi}\sqrt{\frac{S}{\pi}}\Big).
\end{array}
\end{equation}

In the absence of the quantum effects but considering the presence of quintessence dark energy ($\beta=0$, $a\neq 0$), the temperature of Eq.(\ref{th6}) becomes
\begin{equation}\label{th7}
T_{H}=\frac{1}{4(\pi S)^{\frac{3}{2}}}\Big(\pi S+3a\omega_{q}\pi^{\frac{3\omega_{q}+3}{2}}S^{\frac{1-3\omega_{q}}{2}}\Big).
\end{equation}
This is the expression of Hawking temperature for a Schwarzschild black hole surrounded by quintessence \cite{Ri20}. \\In the absence of quintessence $(a=0)$ and for $\beta=0$, the expression of  Eq.(\ref{th6}) is reduced to

\begin{equation}\label{th8}
     T_{H}=\frac{1}{4\sqrt{\pi S}}=\frac{1}{4\pi r_{h}}=\frac{1}{8\pi m},
\end{equation}
which is identical to the expression obtained for the Schwarzschild black hole free from any external field \cite{Rt8}.

\begin{figure}[h]
  \begin{center}
  \includegraphics[width=8cm]{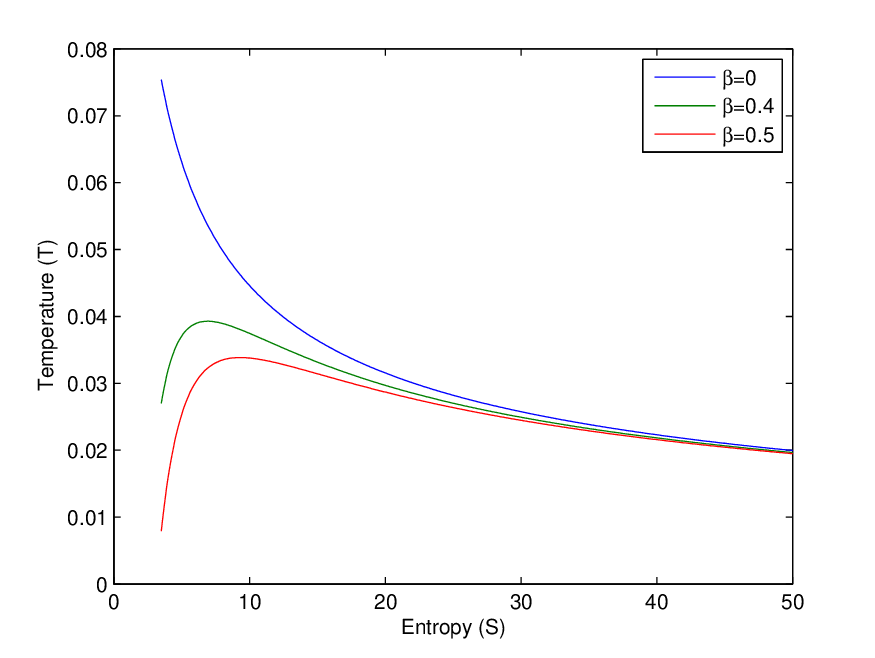}\\
  \caption{Behavior of the Hawking temperature versus the entropy for different values $\beta$, with $a=0$.}\label{mm011}
  \end{center}
\end{figure}

\begin{figure}[h]
  \begin{center}
  \includegraphics[width=8cm]{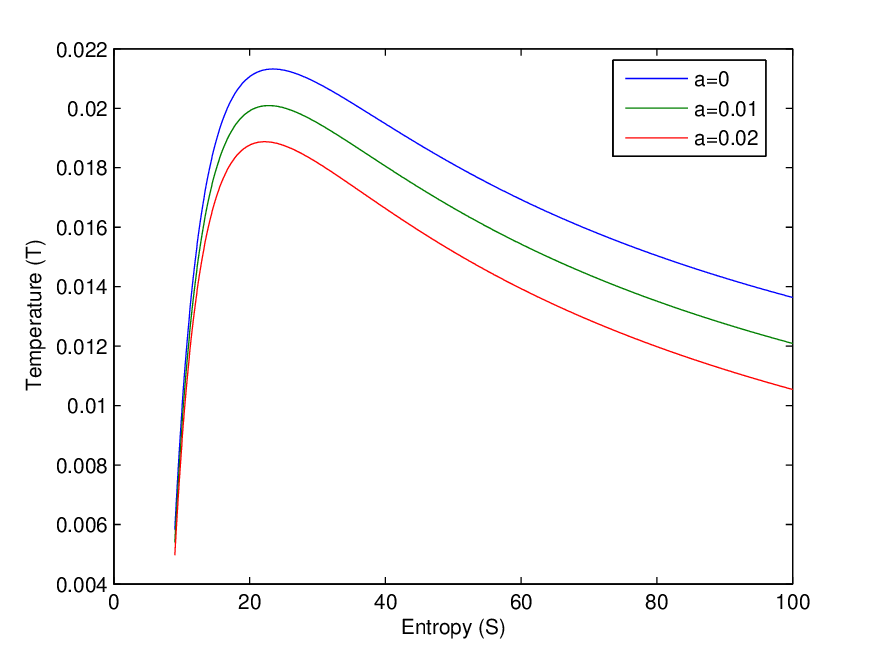}\\
  \caption{Behavior of the Hawking temperature versus the entropy for different values of the quintessence normalization factor, $\omega_{q}=-2/3$, $\beta=1$ .}\label{mm1}
  \end{center}
\end{figure}

The behavior of the Hawking temperature is explicitly plotted on Fig.\ref{mm011} and Fig.\ref{mm1}.

Fig.\ref{mm011} represents the behavior of Hawking temperature of the black hole in the absence of the quintessence $(a=0)$. Through the figure, we can see that taking account the perturbation due to the quantum effects (for $\beta=0$), the Hawking temperature decreases when the entropy increases. This is the behavior of the Hawking temperature for a Schwarzschild black hole. Setting $\beta \neq 0$, the temperature increases and reaches a peak, where the evaporation of black hole is maximum before decreasing \cite{R1201}. Moreover, when increasing the values of $\beta$, the Hawking temperature of black hole decreases.

In Fig.\ref{mm1}, we can see that for a fixed value of $\beta\neq 0$, the increase of the values of quintessence normalization factor $a$ induces a decrease of the Hawking temperature of the regular Hayward black hole surrounded by quintessence. It results that the quintessence accentuates the cooling of the regular Hayward black hole.

Furthermore, we see that for low values of the entropy, the temperature increases. When we pass to the higher entropy, the temperature starts to decrease with entropy.

Thus from the different figures (Fig.\ref{mm011} and Fig.\ref{mm1}) we can understand that, the presence of the quantum effects induces a cooling of the black hole which is accentuated by the presence of quintessence dark energy.

Since the entropy of the black hole $S=\pi r_{h}^{2}$ depends on the horizon size which depends on the the black hole mass $m$, the study the black hole thermodynamics can be made through the behaviors of the above quantities when varying $r_{h}$ or $m$. In Refs. \cite{Rik20,Rik21} that is done using $m'$. Comparing their results with ours, we can see that these thermodynamics quantities have the same behaviors as $m$ and $S$ evolve in the same way.

Moreover, using the equations Eq.(\ref{th3}), Eq.(\ref{th5}) in Eq.(\ref{eqcor1jmpa02}) one can obtains the thermodynamic potential induce by the quintessence under form
\begin{equation}\label{eqcor1jmpa04}
V_{a}=-\frac{1}{2}\big(\frac{S}{\pi}\big)^{-\frac{3\omega_{q}}{2}}.
\end{equation}.

Furthermore, substituting Eq.(\ref{th3}), Eq.(\ref{th5}) in Eq.(\ref{eqcor1jmpa03}), the thermodynamic potential dues to the quantum effects can be expressed as
\begin{equation}\label{eqcor1jmpa05}
V_{\beta}=2\beta\frac{\big(\frac{S}{\pi}\big)^{\frac{3\omega_{q}+1}{2}}-a}{\big(\frac{S}{\pi}\big)^{(\frac{3\omega_{q}}{2})}\Big(\big(\frac{S}{\pi}\big)^{(\frac{3}{2})}+2\beta^{2}\Big)}
\end{equation}.

These two expression (Eq.(\ref{eqcor1jmpa04}) and Eq.(\ref{eqcor1jmpa05})) show that, the thermodynamic potential $V_{\beta}$  depends on both parameter $\beta$ due to the quantum effects and to the parameters $a$ and $\omega_{q}$ due to the quintessence dark energy candidate. However, the potential $V_{a}$ depends only to the state parameter of quintessence dark energy $\omega_{q}$.
\section{Specific heat and phase transition of the Hayward black hole with quintessence}
\label{sec:d}
Knowing the expression of the black hole temperature, the specific heat is deduced by the following relation \cite {R4,Rt7,Ri20}
\begin{equation}\label{sp1}
    C=T_{h}\Big(\frac{\partial S}{\partial T_{h}}\Big).
\end{equation}
Substituting Eq.(\ref{th6}) in Eq.(\ref{sp1}), the specific heat can then be expressed as

\begin{equation}\label{sp2}
    C=-2S^{\frac{3}{2}}\big(S^{\frac{3}{2}}+2\pi\sqrt{\pi}\beta^{2}\big)\frac{H(S,\beta)}{D(S,\beta)},
\end{equation}
where
\begin{equation}\label{sp02}
\begin{array}{ll}
    H(S,\beta)=& 3a\big(2\pi^{2}\beta^{2}(\omega_{q}+1)+\omega_{q}S\sqrt{\pi S}\big)\big(\frac{\pi}{S}\big)^{\frac{3\omega_{q}}{2}}\\
    & -4\pi\beta^{2}\sqrt{\pi S}+S^{2},
\end{array}
\end{equation}
and
\begin{equation}\label{sp002}
\begin{array}{ll}
    D(S,\beta)=& S^{4}-4\pi S\beta^{2}\big(2\pi\beta^{2}+5\pi^{\frac{1}{2}}S^{\frac{3}{2}}\big)
+a\Big(\big(3\omega_{q}S^{4}(3\omega_{q}+2)\\
&+36\pi^{3}S\beta^{4}(\omega_{q}+1)(\omega_{q}+\frac{}2{3})\big)\big(\frac{\pi}{S}\big)^{\frac{3\omega_{q}+1}{2}}\\
 &  +6\pi^{2}S^{2}\beta^{2}\big(6\omega^{2}_{q}+7\omega_{q}+5\big)\big(\frac{\pi}{S}\big)^{\frac{3\omega_{q}}{2}}\Big).
 \end{array}
\end{equation}

Setting the parameter $\beta=0$, for $a\neq 0$,  the expression of the specific heat is reduced to
\begin{equation}\label{sp3}
    C=-2S\frac{1+3a\omega_{q}\big(\frac{\pi}{S}\big)^{\frac{3\omega_{q}+1}{2}}}{1+3a\omega_{q}\big(3\omega_{q}+2\big)
    \big(\frac{\pi}{S}\big)^{\frac{3\omega_{q}+1}{2}}}
\end{equation}
 This expression of Eq.(\ref{sp3}) represents the specific heat for the Schwarzschild black hole surrounded by
quintessence \cite{Ri20}.

 For $a=0$ and $\beta\neq 0$, the expression of the specific heat takes the form
 \begin{equation}\label{sp4}
    C=-2S\frac{\big(S^{2}+2\pi\beta^{2}\sqrt{\pi S}\big)\big(S^{2}-4\pi\beta^{2}\sqrt{\pi S}\big)}{S^{4}-4\pi S\beta^{2}\big(2\pi^{2}\beta^{2}+5S\sqrt{\pi S}\big)},
 \end{equation}
which is the expression of the specific heat for the regular Hayward black hole.

Setting $\beta=0$, Eq.(\ref{sp4}) the specific heat is transformed  to $C=-2S$, which corresponds to the specific heat of the Schwarzschild black hole free from any kind of perturbation \cite{R4,Rt8,Ri20}. The specific heat is negative showing that the black hole is thermodynamically unstable \cite{R4,Rt7}.

\begin{figure}[h]
  \begin{center}
  \includegraphics[width=8cm]{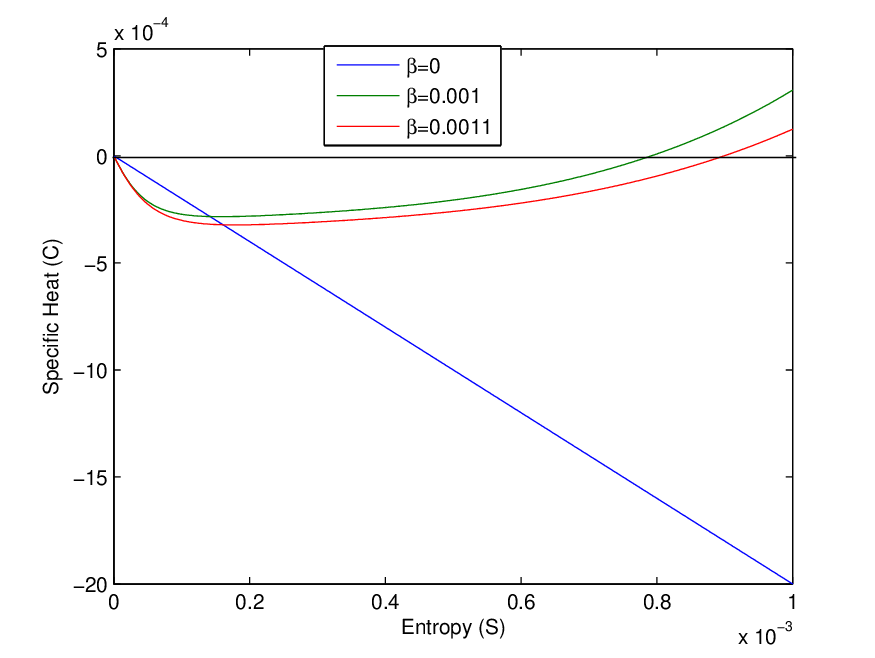}\\
  \caption{Behavior of the specific heat versus the entropy (lower entropies) for different values of $\beta$, with $\omega_{q}=0$, $a=0$ .}\label{mm00}
  \end{center}
\end{figure}

\begin{figure}[h]
  \begin{center}
  \includegraphics[width=8cm]{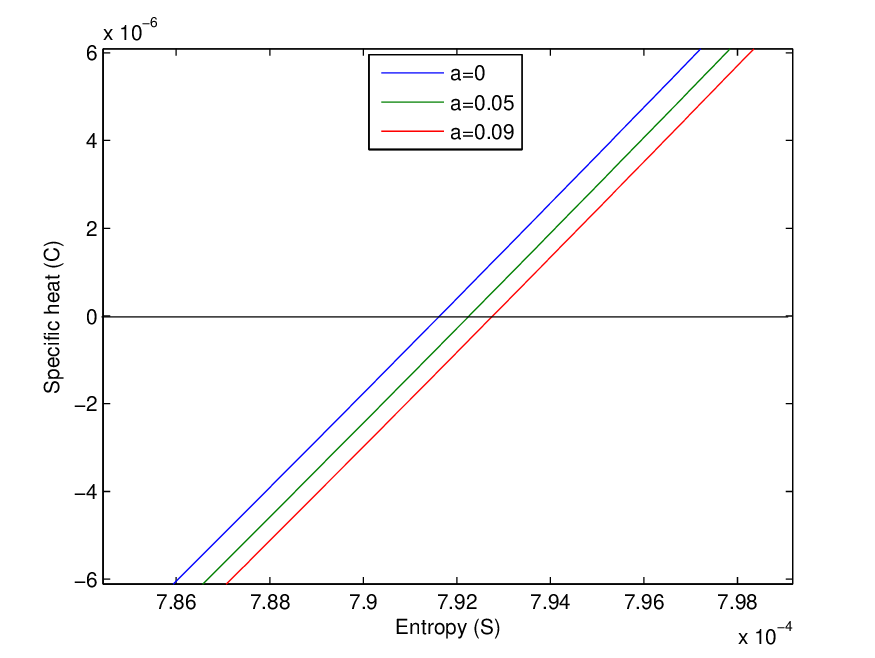}\\
  \caption{Behavior of the specific heat versus the entropy (lower entropies) for different values of the quintessence normalization factor $(a)$, with $\omega_{q}=-2/3$, $\beta=0.001$. }\label{mm01}
  \end{center}
\end{figure}

\begin{figure}[h]
  \begin{center}
  \includegraphics[width=8cm]{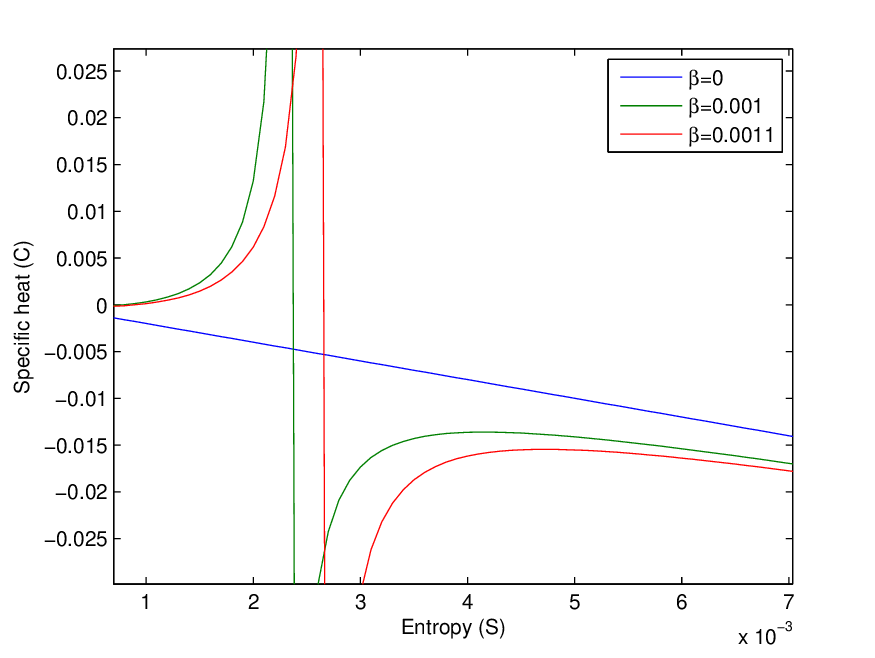}\\
  \caption{Behavior of the specific heat versus the entropy (higher entropies) for different values of $\beta$, with $\omega_{q}=-2/3$, $a=0$.}\label{mm0}
  \end{center}
\end{figure}

\begin{figure}[h]
  \begin{center}
  \includegraphics[width=8cm]{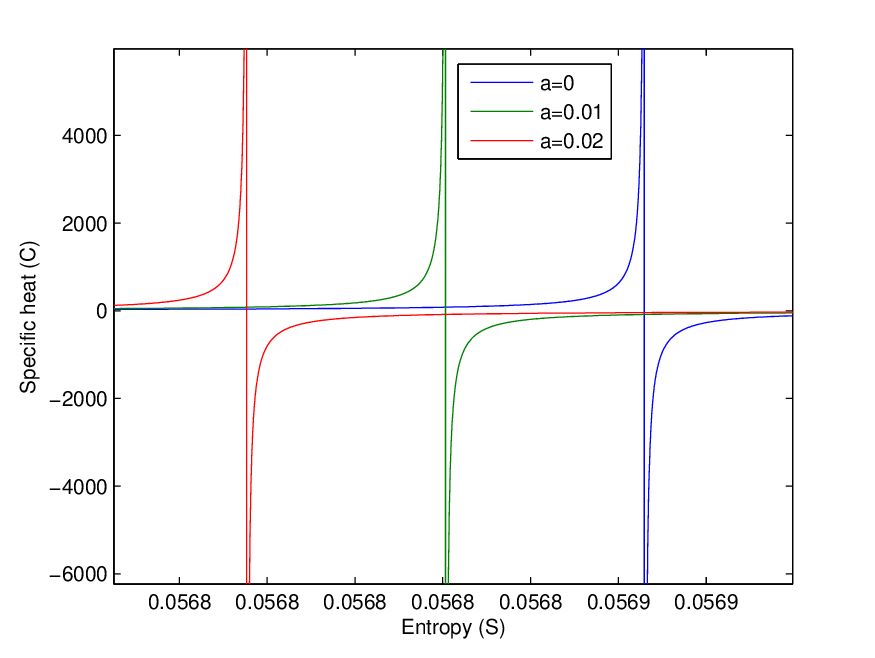}\\
  \caption{Behavior of the specific heat versus the entropy (higher entropies) for different values of the quintessence normalization factor $(a)$, with $\omega_{q}=-2/3$, $\beta=0.011$.}\label{mm}
  \end{center}
\end{figure}

 We explicitly plot the behavior of the specific heat for different values of $\beta$ and the quintessence normalization factor  $a$ versus the black hole entropy. This is shown in Fig.\ref{mm00}, Fig.\ref{mm01}, Fig.\ref{mm0} and Fig.\ref{mm}.

 In Fig.\ref{mm00} we see that, considering the quantum effects ($\beta\neq 0$), for the lower entropies, it occurs a continuity point where the specific heat graph passes from negative to positive values. This shows that it appears a first order thermodynamics phase transition in the black hole. It means that, the black hole passes from unstable to stable phase.

Fig.\ref{mm01} represents the behavior of the specific heat versus the entropy for a fixed value of the magnitude of correction term due to the quantum effects, and assuming the presence of quintessence, for the lower entropies. It results that, when increasing the magnitude of the quintessence normalization factor, the the first order thermodynamics phase transition point is shifted to the higher entropies.

We remark in Fig.\ref{mm0} that, for the higher entropies (large black holes), the parameter $\beta$ induces a discontinuity point where the specific heat-entropy graph passes from positive values to negative values. We see that the black hole passes from stable phase to unstable phase. Moreover, at that last phase, the specific heat remains low compared to that of the Schwarzschild black hole. It results that the regular Hayward black hole is more unstable than the Schwarzschild black hole for the higher entropies. Furthermore, when increasing the magnitude of $\beta$, the first and second order phase transition are shifted to the higher entropies.

 Fig.\ref{mm} shows that, for a fixed value of parameter $\beta \neq 0$, setting $a\neq 0$, the discontinuity point is always present. Moreover, we see that when increasing of the value of the quintessence normalization factor, the discontinuity point is shifted to lower entropies.

Through theses two figures ( Fig.\ref{mm01} and Fig.\ref{mm}), it appears that the quintessence accentuates the instability of the regular Hayward black hole for a fixed value of $\beta \neq 0$.

However, to confirm the stability of a physical system, we need to prove that its potential energy presents a minimum. The only thermodynamics quantity to investigate this behavior is the Gibbs free energy.

%
\section{ Gibbs free energy and global thermodynamic stability analysis of the Hayward black hole with quintessence}
\label{sec:cg}
Gibbs free energy or simply Gibbs energy is the thermodynamics quantity introduced by the physicist Willard Gibbs to analyze the global behavior of a thermodynamic system. Indeed using the classical definition, the Gibbs free energy is defined by \cite{gib01}
\begin{equation}\label{Gib1}
    G=H-TS=U+PV-TS
\end{equation}
Where $G$ is the Gibbs free energy, $H=U+PV$ is the enthalpy of the thermodynamic system, $P$ is the pressure induces on the black hole by the quintessence
, $V=\frac{4}{3}\pi (\frac{S}{\pi})^{\frac{3}{2}}=\frac{4}{3}S\sqrt{\frac{S}{\pi}}$ is the volume of the black hole surface, $U$ the black hole internal energy, $T$ is the temperature of the system and $S$ is the entropy of the system.
Taking into account Kiselev's investigations on the solution to Einstein's equations describing a black hole surrounded by quintessence \cite{Re9,Re10,Re11}, the pressure $P$ that quintessence dark energy induces on the black hole system is connected to the density of quintessence $\rho_{q}$ by the equation of state under the form
\begin{equation}\label{gs1}
P=\rho_{q}\omega_{q}
\end{equation}
Where $\rho_{q}=-\frac{a}{2}\frac{3\omega_{q}}{r^{3(\omega_{q}+1)}}$ and $-1 \leq \omega_{q} \leq -\frac{1}{3}$. Thus the pressure can be written in the form \cite{crf1}
 \begin{equation}\label{Gib01}
P=-\frac{a}{2}\frac{3\omega^{2}_{q}}{r^{3(\omega_{q}+1)}}
\end{equation}
Then the Gibbs free energy becomes \begin{equation}\label{Gib4}
    G=M-2a\frac{\omega^{2}_{q}}{r^{3(\omega_{q}+1)}}S\sqrt{\frac{S}{\pi}}-TS
\end{equation}
Substituting the Eq.(\ref{th3}), Eq.(\ref{th6}) in Eq.(\ref{Gib4}), the Gibbs free energy for Hayward black hole surrounded by quintessence can be expressed as
\begin{equation}\label{Gib5}
\begin{array}{ll}
    G=&\frac{\pi}{2}\frac{\Big((\frac{S}{\pi})^{\frac{(3\omega_{q}+1)}{2}}-a\Big)
    \Big((\frac{S}{\pi})^{\frac{3}{2}}+2\beta^{2}\Big)}{2S(\frac{S}{\pi})^{\frac{(3\omega_{q}+1)}{2}}}-2a\frac{\omega^{2}_{q}}{r^{3(\omega_{q}+1)}}S\sqrt{\frac{S}{\pi}}- \frac{1}{4\big(S\sqrt{\frac{S}{\pi}}+2\pi\beta^2\big)}\times\\
    &  \Big(3a\big(\omega_{q}S\sqrt{\frac{S}{\pi}}+2\pi\omega_{q}\beta^2+2\pi\beta^2\big)\big(\frac{S}{\pi}\big)^{-\frac{3\omega_{q}}{2}} +\frac{S^2-4\pi^2\beta^2}{\pi}\sqrt{\frac{S}{\pi}}\Big).
   \end{array}
\end{equation}

\begin{figure}[h]
  \begin{center}
  \includegraphics[width=8cm]{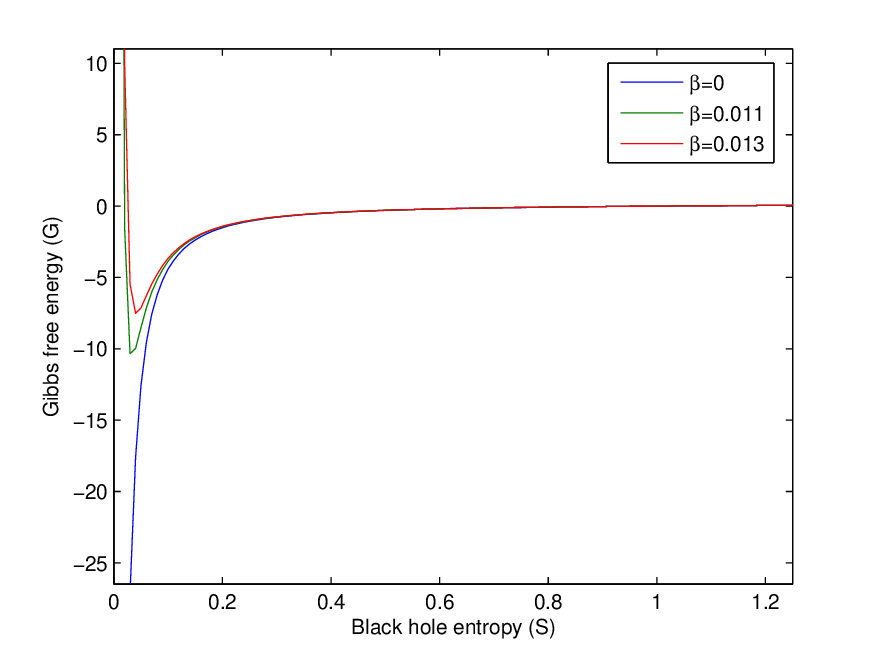}\\
  \caption{Behavior of Gibbs free energy for different values of $\beta$, with $\omega_{q}=-2/3$, $a=0$.}\label{mmk0}
  \end{center}
\end{figure}

\begin{figure}[h]
  \begin{center}
  \includegraphics[width=8cm]{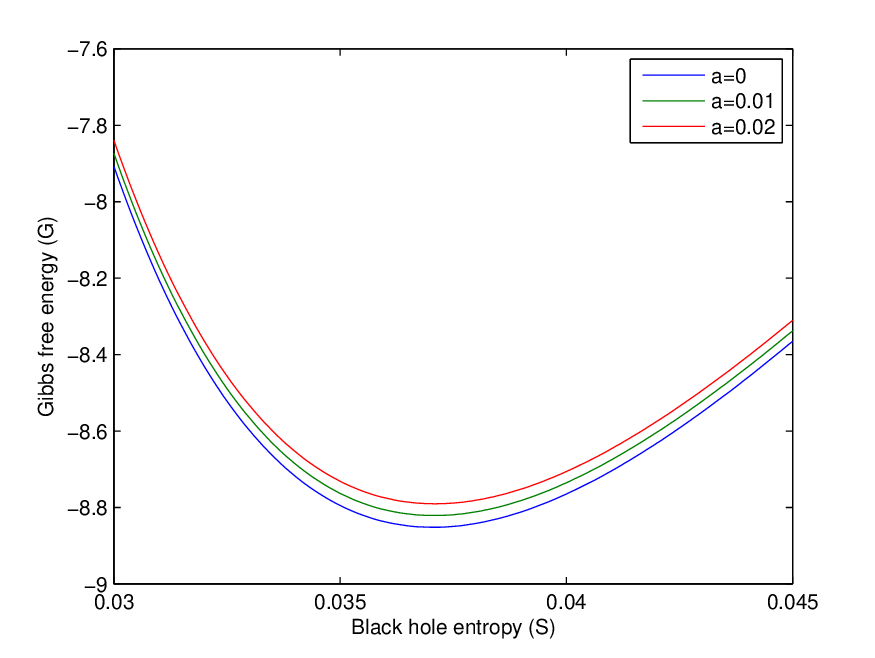}\\
  \caption{Behavior of the Gibbs free energy versus the entropy for different values of the quintessence normalization factor $(a)$, with $\omega_{q}=-2/3$, $\beta=0.012$.}\label{mmk1}
  \end{center}
\end{figure}

The behavior of the Gibbs free energy can be observed on the figures ( Fig.\ref{mmk0} and Fig.\ref{mmk1}) below

Inspecting the Fig.\ref{mmk0}, we remark that, for $\beta=0$ the Gibbs free energy is only growing, thus we can conclude that our system will stay totaly in the unstable phase. However, for $\beta\neq 0$ we observe that, Gibbs free energy starts decreasing before growing again through a minimum, which is characterized as a point where our system can be considered as stable. Moreover, we see that, the minimum point of the Gibbs free energy point is located between the first and second order phase transitions points of Fig.\ref{mm00} and Fig.\ref{mm0}. Thus we can conclude that, for $\beta=0$ our system passes from the unstable phase to stable one by a first order transition, then goes back to the unstable phase by a second order phase transition. Furthermore, the increasing of the magnitude of $\beta$ induces an increase of the Gibbs free energy, thus facilitates the second order phase transition.

Moreover, taking into account the presence of quintessence dark energy $a\neq 0$ and $\beta=0$, we can see through the Fig.\ref{mmk1} that, the increase of the quintessence normalization factor induces also an increase of the Gibbs free energy quantity.

%
\section{Conclusion}
\label{sec:e}
In this work, thermodynamics behavior of the regular Hayward black hole for a larger length scales surrounded by quintessence has been investigated. Using the metric of  the regular Hayward Black hole surrounded by quintessence (Eq.(\ref{em8}) and Eq.(\ref{em9})), associated to the new approach of holographic principle proposed by Verlinde, we have expressed the Unruh Verlinde temperature. Then, using the first law of black hole thermodynamics, we have derived the Hawking temperature and specific heat for a regular Hayward black hole surrounded by quintessence. The Gibbs free energy was also derived using the classical definition of the enthalpy and the expression of pressure defined by the equation of state describing the dark energy quintessence. The behavior of Hawking temperature has been plotted. It results show that when increasing the quintessence parameter, the temperature versus the black hole entropy is decreasing. The behavior of the specific heat versus the entropy has been plotted for different values of $\beta$ (see Fig.\ref{mm00} and Fig.\ref{mm0}). We see that, the parameter $\beta$ is responsible for the first and second-order thermodynamic phase transition in black hole. Increasing the magnitude of $\beta$ moves the phase transition points to the higher entropies. Through Fig.\ref{mm01} and Fig.\ref{mm}, it results that when increasing the magnitude of quintessence normalization factor, the thermodynamics phase transition points are shifted to higher and lower entropies for the first and second order thermodynamics phase transition points respectively. The plot of the Gibbs free energy versus the black hole entropy highlights the existence of a stability zone for the black hole induces by the presence of $\beta$ and confirms the effects of the parameters $\beta$ and $a$ on the black hole thermodynamics phase transition behaviors.
The quintessence accentuates  the instability of regular Hayward black hole, even for the higher magnitude of $\beta$.

\end{document}